\documentclass[useAMS,usenatbib]{mn2e}
\usepackage{graphicx}
\usepackage{amsmath}
\usepackage{color}
\usepackage{amssymb}

\newcommand{\Line}[3]{\Ion{#1}{#2}~#3\,\AA}

\newcommand{\Ion}[2]{#1{\,\sc#2}}
\newcommand{\Rwd}{\mbox{$R_{\mathrm{wd}}$}}
\newcommand{\Mwd}{\mbox{$M_{\mathrm{wd}}$}}
\newcommand{\Msun}{\mbox{$\mathrm{M}_{\odot}$}}
\newcommand{\Rsun}{\mbox{$\mathrm{R}_{\odot}$}}

\begin{document}

\title{Variable Emission from a Gaseous Disc around a Metal-Polluted White Dwarf}
\author[Wilson et
  al. 2014]{D.J. Wilson$^1$\thanks{D.J.Wilson.1@Warwick.ac.uk},
  B.T. G{\"a}nsicke$^1$, D. Koester$^2$, R. Raddi$^1$, E. Breedt$^1$, \newauthor
  J. Southworth$^3$, S.G. Parsons$^4$  \medskip\\
$^{1}$ Department of Physics, University of Warwick, Coventry CV4 7AL,
UK\\
$^{2}$ Institut f\"ur Theoretische Physik und Astrophysik, University of Kiel,
24098 Kiel, Germany\\
$^3$ Astrophysics Group, Keele University,
Staffordshire, ST5 5BG, UK \\
$^4$ Departmento de F\'sica y Astronom\'a, Universidad de Valpara\'so,
Avenida Gran Bretana 1111, Valpara\'iso 2360102, Chile \\
} \date{\today}
\maketitle

\begin{abstract}
We present the discovery of strongly variable emission lines from a
gaseous disc around the DA white dwarf SDSS\,J1617+1620, a star
previously found to have an infrared excess indicative of a dusty
debris disc formed by the tidal disruption of a rocky planetary
body. Time-series spectroscopy obtained during the period 2006-2014
has shown the appearance of strong double-peaked \Ion{Ca}{ii} emission
lines in 2008. The lines were weak, at best, during earlier
observations, and monotonically faded through the remainder of our
monitoring. Our observations represent unambiguous evidence for
short-term variability in the debris environment of evolved planetary
systems. Possible explanations for this extraordinary variability
include the impact onto the dusty disc of either a single small rocky
planetesimal, or of material from a highly eccentric debris tail. The
increase in flux from the emission lines is sufficient that similar
events could be detected in the broadband photometry of ongoing and
future large-area time domain surveys.
\end{abstract}


\section{Introduction}
In 1987 an infrared (IR) excess was discovered around the metal-polluted white dwarf GD29-38 \citep{zuckerman+becklin87-1}. Initially
thought to have been evidence of a brown dwarf companion, it was later
shown by \cite{grahametal90-1} to be emission from a dusty disc in
orbit around the white dwarf, formed by the tidal disruption of an
asteroid \citep{jura03-1}.  Since then similar IR excesses have been
detected around $\simeq$~30 more white dwarfs, and current estimates
suggest that 1--3\% of white dwarfs posses dusty debris disks
\citep{farihietal09-1, girvenetal11-1, steeleetal11-1}. Accretion from
the debris discs results in metal pollution of the white dwarf
atmospheres \citep{koesteretal97-1}, opening up a window on the bulk
abundances of exo-planetary material \citep{zuckermanetal07-1,
  kleinetal10-1, juraetal12-1, gaensickeetal12-1, dufouretal12-1,
  xuetal13-1, barstowetal14-1}. Interestingly, the fraction of white
dwarfs exhibiting traces of metals in their atmospheres is $25-50$\%
\citep{zuckermanetal03-1, zuckermanetal10-1, koesteretal14-1},
suggesting that circumstellar debris too tenuous for detection with
current instrumentation is very frequent \cite[see also][]{bergforsetal14-1}. 

In addition to circumstellar dust, gaseous metallic discs have been
found around a handful of metal-polluted white dwarfs through the
detection of emission lines of the 8600\AA\ \Ion{Ca}{ii} triplet
\citep{gaensickeetal06-3, gaensickeetal07-1, gaensickeetal08-1,
  gaensicke11-1, dufouretal12-1, melisetal12-1, farihietal12-1}. The
double-peaked morphology of the \Ion{Ca}{ii} lines dynamically
constrains the the gas to be located within the Roche radius of the
white dwarf, with the inner disc radii being a few ten white dwarf
radii, broadly consistent with the sublimation radius (but see
\citealt{rafikov+garmilla12-1}). \cite{brinkworthetal09-1,
  brinkworthetal12-1} and \citet{melisetal10-1} showed that the
gaseous and dusty disc components overlap largely in their radial
extension, which is intriguing as the temperature beyond the
sublimation radius should not be sufficiently high to produce the
gaseous disc \citep{hartmannetal11-1}. In addition to this we do not
observe gaseous discs at all white dwarfs with IR excess
\citep{farihietal12-1}, which suggests that the production mechanism
of the gaseous discs is not universal. Plausible scenarios for the
generation of gas are collisions of multiple asteroids
\citep{jura08-1} and sublimation that may lead to a runaway evolution
of the debris discs \citep{rafikov11-2, metzgeretal12-1}.

All the above suggests that the detection of gaseous discs around
metal-polluted white dwarfs may point to a highly dynamic debris
environment. So far, evidence for short-term observational variability
of debris discs is very limited. \citet{gaensickeetal08-1} reported
changes in the line profile shapes of the \Ion{Ca}{ii} emission lines
detected in SDSS\,J084539.17+225728.0, and
\cite{vonhippel+thompson07-1} claimed variability in the strength of
the Ca~K 3934\AA\ absorption line in the prototype GD\,29-38, though
the latter result was disputed \cite{debes+lopez-morales08-1}. The strongest evidence for  variability previously observed in such systems is the drop in the infrared luminosity of the dusty debris disc around WD\,J0959-0200 in 2010 \citep{xuetal14-2}.

Here we report the discovery of a strongly variable gaseous disc
around the metal-polluted white dwarf SDSS\,J161717.04+162022.4. 

\begin{figure*}
    \centering
    \includegraphics[width=18cm]{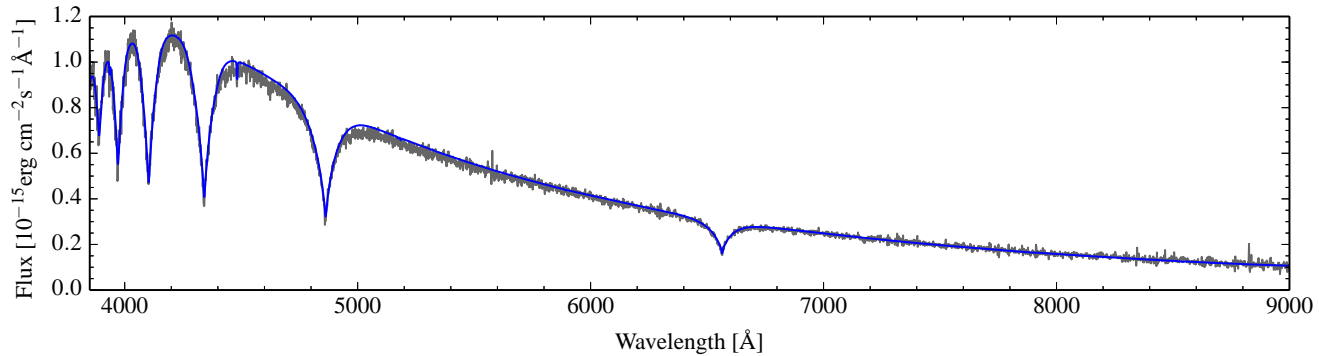}
    \caption{Spectrum of SDSS\,J1617+1620 obtained by SDSS in 2006
      (grey). The atmospheric parameters,
      $T_{\mathrm{eff}}=13520\pm200$\,K and $\log g=8.11\pm0.08$, were
      determined by averaging the fits to both SDSS spectra. The
      corresponding model spectrum is shown in blue. Photospheric
      absorption of Ca~K 3934\,\AA\  and \Line{Mg}{ii}{4481} is
      detected in both SDSS spectra. \protect\label{fig:specfit}}
\end{figure*}

\section{Discovery}
We applied the procedure outlined in \citet{gaensickeetal08-1} to
search for white dwarfs with \Ion{Ca}{ii} emission in Data Release
(DR)~7 of the Sloan Digital Sky Survey (SDSS,
\citealt{abazajianetal09-1}), and identified two new systems, the
He-atmoshere (DB) SDSS\,J073842.56+183509.6 (see also
\citealt{dufouretal12-1}) and the H-atmosphere (DA)
SDSS\,J161717.04+162022.4 (henceforth SDSS\,J1617+1620,
\citealt{gaensicke11-1}). The SDSS spectrum revealed
\Line{Mg}{II}{4481} absorption in the white dwarf atmosphere, and
\textit{Spitzer} observations of SDSS\,J1617+1620 confirmed the
expected presence of circumstellar dust \citep{brinkworthetal12-1}.

The DR\,7 spectrum of SDSS\,J1617+1620 was obtained on 2008 March 3.
Only later, we realised that the star had an earlier DR~6
\citep{adelman-mccarthyetal08-1} spectrum, obtained on 2006 July 1~--~in which the \Ion{Ca}{ii} emission lines are weakly detected, at
best. Intrigued by this clear evidence for variability, we initiated spectroscopic monitoring, using the William Herschel Telescope (WHT),
Gemini North, and the ESO Very Large Telescope (VLT).

\section{Follow-up observations}
The log of the observations is given in Table\,\ref{tab:obs}.
The first follow-up spectra of SDSS\,J1617+1620 were obtained using the Intermediate dispersion Spectrograph and Imaging System (ISIS) spectrograph mounted on the
WHT, with observations carried out on 2009 February 17 and 2010 April 23. On both occasions two 1800s exposures were taken, with a third 540s exposure obtained in 2010.       
Observations were
obtained simultaneously through the blue and red arms of the
instrument. The CCDs were binned by factors of 2 (spectral) and 3
(spatial) to limit the impact of readout noise on the
observations. The blue arm was equipped with the R600B grating and had
a wavelength coverage of 3690--4110\,\AA\ at a reciprocal dispersion
of 0.45\,\AA\ per binned pixel and a resolution of approximately
1\,\AA. The red arm had the R316R grating in 2009 , covering
5760--8890\,\AA\ with a reciprocal dispersion of 1.85\,\AA\ per binned
pixel and a resolution of approximately 3.5\,\AA, and the R1200R grating in 2010, with a reciprocal dispersion of 1.52\,\AA\ per binned
pixel and a resolution of approximately 1.1\,\AA. The data were
reduced and the spectra optimally extracted using the {\sc
  pamela}\footnote{{\sc pamela} and {\sc molly} were written by
  T.\ R.\ Marsh and can be obtained from {\tt
    http://www.warwick.ac.uk/go/trmarsh}} code \citep{marsh89-1} and
the Starlink\footnote{The Starlink software and documentation can be
  obtained from {\tt http://starlink.jach.hawaii.edu/}} packages {\sc
  figaro} and {\sc kappa}. Copper-neon and copper-argon arc lamp
exposures were taken before and after the observations and the
wavelength calibrations were linearly interpolated from them. We
removed the telluric lines and flux-calibrated the target spectra
using observations of BD\,+75\,325 in 2007 and SP1036+433 in 2010.

The next observations were obtained on 2010 June 10 at the Gemini North Telescope, where four 900s exposures were obtained in the i band using the R831 grating, with a central wavelength setting of 8600\AA. The CCDs were binned by a factor of 2x2 and the wavelength was        calibrated using CuAr arcs taken at the end of the night. As with the WHT images, the spectra were again reduced using standard {\sc starlink} procedures, and then optimally extracted using {\sc pamela}. The wavelength calibration was done using {\sc molly}.

\indent Observations at the VLT with X-shooter \citep{vernetetal11-1} in 2011 March--June and the Ultraviolet and Visual Echelle Spectrograph (UVES, \citealt{Dekkeretal00-1})
on 2013 May 05 were reduced using the Reflex\footnote{The Reflex software and documentation can be obtained from {\tt http://www.eso.org/sci/software/reflex/}} software developed by ESO. The
X-shooter data was reduced using the Reflex Nod mode, with the various
parameters of the reduction pipeline varied using trial and error to
produce the best possible spectra. For the UVES data the default
Reflex settings were found to be sufficient. After this a heliocentric
correction was applied to the Reflex products. Where multiple
observations were made on the same night the spectra were combined
using a weighted average.

A final VLT observations was obtained with the FOcal Reducer and low dispersion Spectrograph (FORS, \citealt{appenzelleretal98-2}), taking 3 long-slit
spectra of 300s exposures on 2014 April 30.  We covered the
near-IR spectral range (7750--9250 \AA), using the standard resolution
collimator ($2\times2$), the dispersion grism GRIS\_1028z+29, and the
order separator filter OG590 that give a dispersion of ~0.8\, \AA /pixel,
which corresponds to a resolving power R$\sim$3800 at 8500 \AA. We
reduced the data in a standard fashion, using traditional IRAF
routines for long-slit spectroscopy, i.e.  the 3 spectra were bias
subtracted, flat-fielded, wavelength calibrated, sky-subtracted, flux
calibrated, and finally combined.  The spectrophotometric standard
G138-31 was observed soon after SDSS\,J1617+1620 at similar airmass,
although the night was not photometric and only relative flux
calibration has been possible.

\begin{figure}
    \centering
    \includegraphics[width=8.5cm]{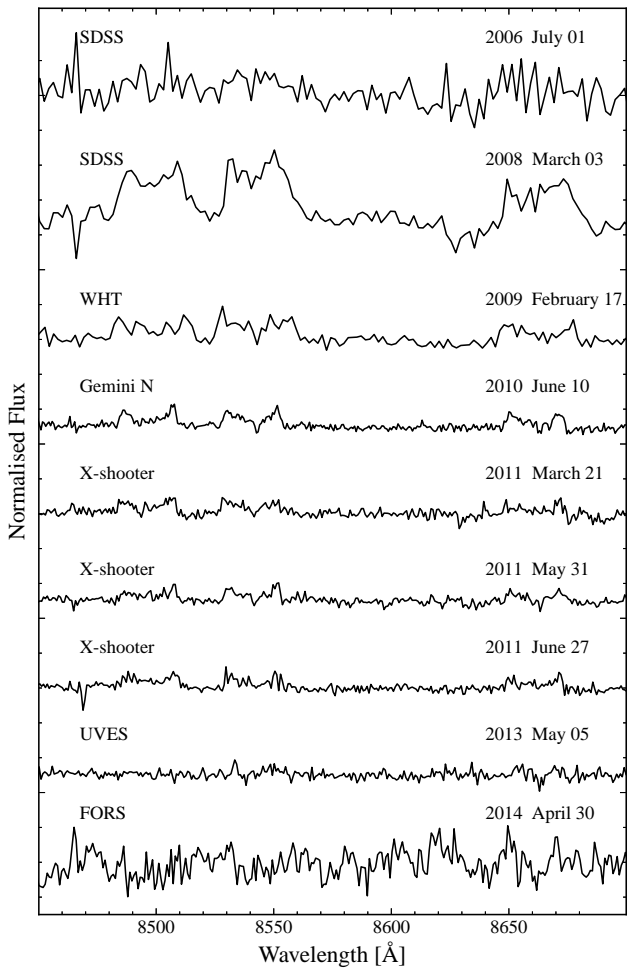}
    \caption{Normalised time-series spectroscopy of SDSS\,J1617+1620
      showing the change in strength of the \Ion{Ca}{ii} 8600\AA\ emission
      line triplet between 2006--2014. The telescope/instrument used to
      make the observation is indicated on the left above each
      spectrum, with the date of the observation on
      the right. On earlier dates the lines clearly show the
      double-peaked morphology characteristic of emission from a
      gaseous disc around the white dwarf
      \citep{horne+marsh86-1}. However by the time that the UVES
      spectrum was obtained in March 2013 the lines, and hence the
      gaseous disc, had disappeared
      (Figure\,\ref{fig:ews}).
       \protect\label{fig:trips}}
\end{figure}

Comparing the two SDSS spectra (Figure\,\ref{fig:trips}) shows the
dramatic increase in the strength of the \Ion{Ca}{ii} 8600\AA\ emission line
triplet between 2006 and 2008, revealing the formation of a
gaseous disc around SDSS\,J1617+1620. The WHT observations, less than a
year later, show a significant reduction in the strength of the
emission lines, as does the Gemini spectra.

The X-shooter spectra show a further decrease in the strength of the \Ion{Ca}{ii} triplet from the WHT and Gemini data, although the emission lines
were still clearly visible on this occasion. However by the time of the
UVES observation in 2013 the emission lines appear to have
disappeared below the detection threshold. Thanks to the high
resolution of the VLT instruments the Ca~K~ 3934\,\AA\ and
\Line{Mg}{ii}{4481} absorption lines are clearly detected and well defined,
allowing for an accurate measurement of the metal abundances in the
atmosphere of SDSS\,J1617+1620.

The FORS spectrum, as with the UVES observations a year before, shows
no indication of the \Ion{Ca}{ii} 8600\AA\ emission lines, confirming the disappearance of the gaseous disc.

\begin{table*}
\centering
\caption{Log of observations SDSS\,J1617+1620}
\begin{tabular}{lcccc}\\
\hline
Date & Telescope/ & Wavelength & Spectral & Total Exposure    \\
& Instrument & Range (\AA) & Resolution (\AA) & Time (s)\\ 
\hline
2006 July 01 & SDSS & 3800-9200 & 0.9 & 5700\\ 
2008 March 03 & SDSS & 3800-9200 & 0.9 & 14700\\
2009 February 17 & WHT & 3690-8890 & 1.6 & 3600\\
2010 April 23 & WHT & 3630-8850 & 0.5 & 3600\\
2010 June 10 & Gemini N & 7750-8960 & 0.7 &3600\\
2011 March 21 & X-shooter & 2990-24790 & 0.2 & 6990 \\
2011 May 31 & X-shooter & 2990-24790 & 0.2 & 16380\\
2011 June 05 & X-shooter & 2990-24790 & 0.2 & 8190\\
2013 May 05 & UVES & 3760-9440 & 0.05 & 4800\\
2014 April 30 & FORS & 7750-9550 & 0.8 & 900\\
\hline 
\end{tabular}
\label{tab:obs}
\end{table*}

\section{White Dwarf Parameters}
\label{sect:wd}
The atmospheric parameters of\,SDSS1617+1620 were calculated by fitting model spectra to the SDSS spectroscopy (Figure\,\ref{fig:specfit})
as described by \cite{koester10-1}. We report the average parameters
obtained from the two SDSS spectra, estimating the uncertainty from
the discrepancy between the two fits, as
$T_{\mathrm{eff}}=13520\pm200$\,K and $\log g=8.11\pm0.08$. The
corresponding mass, radius, and cooling age, computed from the
hydrogen-atmosphere cooling models of Bergeron and
collaborators\footnote{
  http://www.astro.umontreal.ca/$\sim$bergeron/CoolingModels, based on
  \cite{holberg+bergeron06-1, kowalski+saumon06-1,
    tremblayetal11-2}.}, are $\Mwd=0.68\pm0.05\,\Msun$,
$\Rwd=0.0120\pm0.007\,\Rsun$, and $T_{\mathrm{cool}}=350\pm{50}$\,Myr. Using the
initial-mass to final-mass relations of \citet{catalanetal08-2},
\citet{kaliraietal08-1}, \citet{williamsetal09-1} and
\citet{casewelletal09-1} suggests a main-sequence progenitor mass of
$2.2-3.0$\,\Msun, similar to the majority of the metal-polluted white
dwarfs \citep{jura+xu12-1, koesteretal14-1}.

\begin{figure}
    \centering
    \includegraphics[width=8cm]{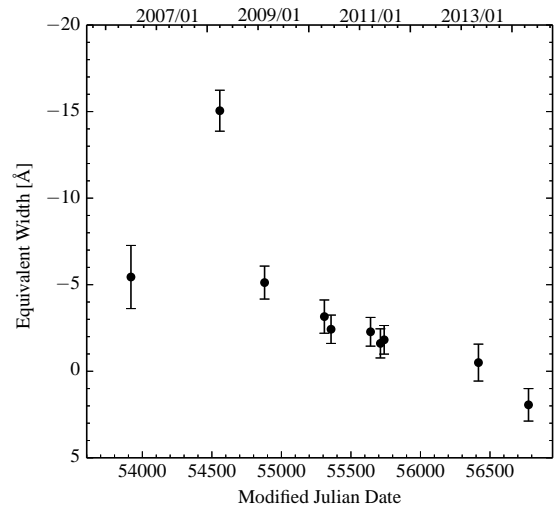}
    \caption{Change in the strength of the \Ion{Ca}{ii} triplet seen in the
      spectra of SDSS\,J1617+1620 over the period 2006--2014 (Figure\,\ref{fig:trips}) . The
      equivalent widths of the emission lines in each spectrum were
      calculated over the wavelength range 8460--8700\AA . The strength
      of the emission line increases by a factor$\sim$3 between the
      first and second observations, before dropping down to zero by
      2013. This represents a dramatic increase then loss of emission from a
      gaseous disc around SDSS\,J1617+1620. \protect\label{fig:ews}}
\end{figure}

\section{Variability of the \Ion{Calcium}{II} Triplet}

The dramatic change in the strength of the \Ion{Ca}{ii} emission lines
is illustrated in Figure \ref{fig:trips}, where the triplet is
clearly seen in the second SDSS spectrum (MJD\,=\,54557), and
subsequently fades during our follow-up spectroscopy. To quantify the
variable nature of the gaseous disc we measure the equivalent widths
of the emission lines. As the 8542 and 8662\,\AA\ components
overlap in some of the spectra, we computed the combined equivalent
width of the entire triplet, using the wavelength range
8460--8700\AA. The resulting values are sensitive to the method used to
fit the underlying continuum, leading to systematic uncertainties that
dominate the error budget in the case of weak or non-existing line
emission. 
The equivalent widths reported in Table\,\ref{tab:widths}, and shown
in Figure\,\ref{fig:ews}, confirm the appearance of strong
\Ion{Ca}{ii} emission lines in 2008, which subsequently faded by a
factor $\simeq2.5$ within less than a year. The decline in the
strengths of the lines then slowed down, and our last spectrum
obtained in April 2014 is consistent with the complete disappearance
of the lines. Unfortunately, the onset of the \Ion{Ca}{ii} emission is
not well-documented. While our analysis formally detects the
\Ion{Ca}{ii} triplet in the 2006, the corresponding SDSS spectrum is
overall of relatively poor quality, and more specifically the
wavelength range relevant for the equivalent measurement is affected
by residuals from the sky line subtraction. We are therefore unable to
unambiguously say whether a gaseous disc was present already in 2006,
or if it only formed in 2008.

Assuming that the dispersion of the gaseous disc by 2014 was due to viscous angular momentum exchange suggests a viscous timescale $t_v\approx$8yr. Using equation 2 of \cite{metzgeretal12-1} yields a viscosity parameter $\alpha\sim$0.25, within the range estimated by \cite{kingetal07-1} for an ionised thin disc.

The \Ion{Ca}{ii} lines show a mildly double-peaked morphology
(Figure\,\ref{fig:trips}), which arises from the Doppler shifts
induced by the Keplerian velocity of the material in the disc
\citep{horne+marsh86-1}. The total width and peak separation
of the \Ion{Ca}{ii} lines can be used to estimate the inner and out
radii of the disc respectively, and we find $R_\mathrm{in}\sin^2
i\approx0.5\,\Rsun$ and $R_\mathrm{out}\sin^2 i\approx1.2\,\Rsun$,
where $i$ is the unknown inclination of the disc. These values are similar
to the inner and outer radius estimated for SDSS\,J122859.93+104032.9
and SDSS\,J084539.17+225728.0 \citep{gaensickeetal06-3,
  gaensickeetal08-1}, i.e. the outer radius is compatible with the
tidal disruption radius of a rocky asteroid \citep{davidsson99-1}, and
the inner radius is near the sublimation radius \citep{vonhippeletal07-1}. Note that, whilst the inner radius can be constrained by calculating the Doppler shift at the Full Width-Zero Intensity (FWZI)  of the emission lines, the point with represents the outer edge is somewhat more arbitrary. In this case we chose to measure the peak separation of the lines to provide a lower limit.  

Inspecting our WHT, Gemini, and
VLT/X-shooter spectra, there is a slight hint that the width of the
\Ion{Ca}{ii} lines decreases with time, which would imply the inner
radius of the gas disc moving further out. However the emission lines
in the later observations are too weak to make a firm
conclusion.

\begin{table*} 
\centering
\caption{Equivalent widths of the absorption and emission lines in the
  times-series spectra of SDSS\,J1617+1620. Note the relative
  consistency of the Ca~K~ 3934\,\AA\ and
\Line{Mg}{ii}{4481} absorption lines
  compared to the hugely variable \Ion{Ca}{ii} 8600\AA\ triplet. No
  measurement was made of the absorption lines in four cases: The SDSS
  spectra are of insufficient quality for an accurate measurement, and the Gemini and
  FORS observations did not cover the corresponding wavelength range.}
\begin{tabular}{lccc}
\hline
& \multicolumn{3}{c}{Equivalent Width [\AA]} \\

Date & \Ion{Ca}{ii} 3934\,\AA\ & \Ion{Mg}{II} 4481\,\AA\   & \Ion{Ca}{ii} 8600\,\AA\ Triplet \\
\hline
2006 July 01 & - & - & -6.4$\pm$1.8\\
2008 March 03 & - & - & -16.1$\pm$1.2\\
2009 February 17 & 0.20$\pm$0.02 & 0.34$\pm$0.02 & -6.1$\pm$1.0\\
2010 April 23 & 0.16$\pm$0.03 & 0.36$\pm$0.05 & -4.2$\pm$1.0\\
2010 June 10 & - & - & -3.4$\pm$0.8\\
2011 March 21 & 0.21$\pm$0.01 & 0.30$\pm$0.02 & -3.3$\pm$0.8\\
2011 May 31 & 0.20$\pm$0.01 & 0.28$\pm$0.01 & 2.-6$\pm$0.8\\
2011 June 27 & 0.20$\pm$0.01 & 0.26$\pm$0.01 & -2.8$\pm$0.8\\
2013 May 05 & 0.23$\pm$0.01 & 0.26$\pm$0.01 & -1.5$\pm$1.1\\
2014 April 30 & - & - & -0.9$\pm$0.9\\
\hline
\end{tabular}

\label{tab:widths}
\end{table*}

\section{Accretion of Planetary Material}
\label{sect:accretion}
\begin{figure}
    \centering
    \includegraphics[width=8cm]{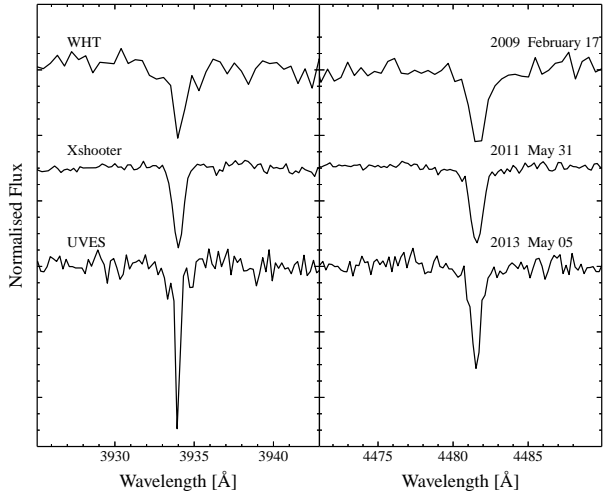}
    \caption{Photospheric absorption lines of Ca~K 3934\,\AA\ (left)
      and \Line{Mg}{ii}{4481} (right), illustrating the different
      spectral resolutions of the instruments used for the follow-up
      spectroscopy. The equivalent widths measured from all our
      spectra are given in Table\,\ref{tab:widths} and average
      accretion fluxes are given in Table\,\ref{tab:abs}. The apparent greater depth of the UVES        observation of the Ca~K line is an effect of the spectral resolution and does not represent an increase in equivalent width.  
      \protect\label{fig:lines}}
\end{figure}

In addition to the \Ion{Ca}{ii} 8600\,\AA\ emission line triplet we detect photospheric absorption of Ca~K~ 3934\AA\ and
\Line{Mg}{ii}{4481} (Figure\, \ref{fig:lines}). The detection of metal pollution provides an
opportunity to investigate the chemical diversity of extrasolar
planetary systems \citep[e.g.][]{zuckermanetal07-1, kleinetal11-1,
  gaensickeetal12-1, xuetal14-1}. The relevant procedures and
detailed physics have been extensively described by
\cite{gaensickeetal12-1} and
\cite{koesteretal14-1}, and we provide here only a brief summary.

A key assumption in the interpretation of the photospheric metal
abundances is a steady state between accretion and diffusion
\citep{koester09-1}, in which case the diffusion flux is constant
throughout the atmosphere, and equal to the accretion rate from the
debris disc. The relativity low effective temperature of
SDSS\,J1617+1620 implies that the diffusion of metals within the
atmosphere of should not be affected by radiative levitation
\citep{chayeretal95-1}. While the temperature is sufficiently high
that no deep convection zone develops, some convection zones are
present at Rosseland optical depths $0.04\la\tau_\mathrm{R}\la3.2$ and
$300\la\tau_\mathrm{R}\la1936$. In between these zones the atmosphere is
radiative. Regardless of the details of the atmospheric
structure, the diffusion time scales are very short
($\tau_\mathrm{diff}\approx0.5$\,yr, Table\,\ref{tab:abs}), meaning that
SDSS\,J1617+1620 must be currently accreting from an external
source, almost certainly the circumstellar gas and dust. We have computed
the corresponding diffusion fluxes under two assumptions: 1. Treating the
entire atmosphere as convective and 2. Evaluating the diffusion time
scales at $\tau_\mathrm{R}=3.2$, the bottom of the shallow
convection zone near the surface. 

Table\,\ref{tab:abs} reports the average accretion fluxes of Mg and
Ca, and an upper limit for Si, obtained from the analysis our WHT,
X-Shooter and UVES spectra (the quality of the SDSS spectra is too
low, and the Gemini and FORS spectra do not cover the relevant
wavelength range). Scaling for the relative abundance of Mg and Ca in
the bulk Earth \citep{allegreetal01-1}, we estimate a total metal
accretion flux onto the white dwarf of
$\approx(6.4\pm1.8$--$7.8\pm3.3)\times10^8\,\mathrm{g\,s^{-1}}$, consistent with
the accretion rates measured in other dusty DA white dwarfs
\citep{vennesetal10-1, vennesetal11-1, melisetal11-1,
  gaensickeetal12-1}. 

We find that the planetary debris around SDSS\,J1617+1620 has
$\log[\mathrm{Ca/Mg}]=-1.4$ to $-1.6$, (by number, depending on the
treatment of convection), which is somewhat low compared to the values
found for the the bulk Earth ($-1.17$, \citealt{mcdonough00-1}) C\,I
chondrites ($-1.24$, \citealt{lodders03-1}), and the Sun ($-1.24$,
\citealt{lodders03-1}). A depletion relative to other elements is also
seen in GALEX\,J193156.8+011745 \citep{vennesetal10-1, vennesetal11-1,
  melisetal11-1, gaensickeetal12-1}, and was interpreted by
\citet{melisetal11-1} as potential evidence for the accretion of a
differentiated body that had its crust and part of its mantle
stripped. Conclusions regarding the nature and origin of the debris
around SDSS\,J1617+1620 remain very limited as only two photospheric
elements are so far detected. Progress will require deep optical and
ultraviolet spectra probing for the abundances of C, Si, Al, Ti, and
Fe.

A final note concerns the temporal evolution of the photospheric
absorption lines of Mg and Ca. While the equivalent widths of the Ca~K
line are constant within the uncertainties (Table\,\ref{tab:widths}),
the \Line{Mg}{ii}{4481} lines shows a notionally significant change,
being strongest in the first WHT observation during the early decline
of the \Ion{Ca}{ii} emission lines. \cite{rafikov11-2} and
\cite{metzgeretal12-1} showed that the generation of gas can lead to a
significant increase in the accretion onto the white dwarf. While the
observations do not exclude small changes in the accretion rate of
Mg (which, if real, would imply a chemical differentiation within the
debris), they do rule out a large variation of the total accretion rate. 

\begin{table}
\centering
\caption{Diffusion timescales and average accretion fluxes, computed
  under the assumption of a fully convective atmosphere or using the
  diffusion time scales at the bottom of the shallow convection zone
  near the surface.}
\begin{tabular}{lrrrr}
\hline
Element &
\multicolumn{2}{c}{fully convective} &
\multicolumn{2}{c}{convective at $\tau_\mathrm{R}=3.2$}  \\
 & 
$\log \tau_\mathrm{diff}$ [yrs] & $\dot m~[\mathrm{g\,s^{-1}}]$ &
$\log \tau_\mathrm{diff}$ [yrs] & $\dot m~[\mathrm{g\,s^{-1}}]$  \\
\hline
12 Mg & $-0.46$ & $1.7\times10^8$       & $-2.2$ & $7.3\times10^7$ \\
14 Si & $-0.36$ & $\leq2.5\times10^8$   & $-2.5$ & $\leq2.9\times10^8$ \\
20 Ca & $-0.37$ & $7.2\times10^6$       & $-2.3$ & $5.0\times10^6$ \\
\hline 
\end{tabular}
\label{tab:abs}
\end{table}

\section{Discussion}

The formation mechanism of gaseous discs around metal-polluted white
dwarfs is very uncertain.  Gaseous discs are only observed around a
small fraction of the metal-polluted white dwarfs, and in all cases
where \Ion{Ca}{ii} emission lines are observed circumstellar dust is
also detected in the form of a noticeable infrared excess
\citep{brinkworthetal09-1, melisetal10-1, melisetal12-1,
  brinkworthetal12-1, farihietal12-1}. However, many dusty and
strongly metal-polluted white dwarfs do not show any emission lines
from circumstellar gas \citep[e.g.][]{gaensickeetal07-1,
  vennesetal10-1, kleinetal11-1, farihietal12-1}, so their formation cannot be universal. Gaseous discs are found around white
dwarfs with temperatures ranging from $\simeq13\,000$\,K to
$\simeq22\,000$\,K \citep{gaensickeetal06-3, gaensickeetal07-1,
  gaensickeetal08-1, farihietal12-1, melisetal12-1,dufouretal12-1},
very similar to the range where dusty discs are found
\citep{farihietal09-1}. 

The \Ion{Ca}{ii} emission lines can be modelled by optically thin H
and He-deficient gas with a temperature near 6000\,K, but the heating
mechanism is unclear: \citet{hartmannetal11-1} assume an active disc,
i.e. heated by the inwards flow of material, but require an
unrealistically high accretion rate of $10^{17-18}\mathrm{g\,s^{-1}}$,
which is many orders of magnitude higher than the accretion rates onto
the white dwarf derived from the photospheric abundances. \citet{kinnear11}
and \citet{melisetal10-1} showed that irradiation from the white dwarf
is sufficient to keep circumstellar metal gas at many 1000\,K, and
that cooling occurs primarily through optically thick lines. \citet{kinnear11} could quantitatively reproduce the observed line
fluxes with a photo-ionisation model.

The high temperature necessary to explain the \Ion{Ca}{ii} emission
lines, together with the radial extent of the gas discs
($\sim0.5-1.5$\,\Rsun) and the absence of gas at many dusty white
dwarfs rules out production of the gas by sublimation of the dust
within the radiation field of the white dwarf. In fact
\citet{gaensickeetal06-3, gaensickeetal07-1} and
\citet{brinkworthetal12-1} showed that the location of the inner edges
of the gas discs derived from observations are broadly consistent with
the sublimation radius, i.e. the radial distribution of gas and dust
largely overlap (for completeness, we note that line-of-sight
absorption of circumstellar gas has been detected in two cases, see
\citealt{debesetal12-2, gaensickeetal12-1}, but the exact location of
that gas is not well known).

We speculate on two possible explanations for the transient formation of the
gaseous disc. One scenario is the impact of a small asteroid onto a
more massive, pre-existing debris disc \citep{jura08-1}. The rocky nature of this body is implied by the absence of Balmer emission lines from the disc. Entering the
tidal disruption radius, the incoming asteroid will start to break up,
and vaporise upon the high-velocity impact onto the disc. The gas
generated in that way will subsequently spread radially due to viscous
angular momentum exchange, and heating by the white dwarf will result
in the observed \Ion{Ca}{ii} emission. The gas will eventually accrete
onto the white dwarf (with a small amount moving outwards to carry
away angular momentum), and the decreasing density will result in a
weakening of the emission lines. In summary, such a ``secondary
impact'' event would be produce a single event of transient
\Ion{Ca}{ii} emission. 

A slight twist to the above scenario is that the dust disc at
SDSS\,J1617+1620 is young, and we have witnessed the impact of
material in a debris tail left-over from the original disruption.  The
detailed evolution of the tidal disruption of asteroids has not yet
been fully explored. \citet{debesetal12-1} simulated the disruption of
a rubble pile asteroid, and found that an elongated debris train is
formed. \citet{verasetal14-1} followed the evolution of a disrupted
rubble piles over many orbital cycles, and demonstrated that, in the
absence of additional forces beyond gravity, a highly collisionless
eccentric ring of debris is formed. Additional forces, e.g. from
sublimation, are probably required to fan out the debris train, eventually
circularising material in a close-in circumstellar disc. This process
is likely to extend over many orbital cycles, which, as the original semi-major axis of the asteroid
 should have been $>1-2$\,au (i.e. beyond the
region cleared out during the red giant phase), could imply timescales of many tens
to even thousands of years. During that period, repeated impacts of
left-over debris are expected, and one would expect another flare-up
of gas emission from SDSS\,J1617+1620 over the next years to decade.

The serendipitous discovery of the transient \Ion{Ca}{ii} emission in
SDSS\,J1617+1620 is a strong motivation for a more systematic
monitoring of large numbers of white dwarfs to detect tidal disruption
events of planetary bodies. The spectroscopic observations that led to
the discovery of the gas disc in SDSS\,J1617+1620 are moderately
expensive in terms of telescope time and aperture. The observed changes in the \Ion{Ca}{ii} line fluxes correspond to a
2\% variability of SDSS\,J1617+1620 in the $z$-band. Such events are
hence close to the detection threshold of current ground-based
transient surveys \citep{izevicetal07-1, ofeketal12-1}, but should be
easily detected in the era of Gaia and LSST \citep{carrascoetal14-1,
  lsst09-1}.

\section{Conclusion}
We have observed unambiguous evidence for a variable gaseous disc
around the white dwarf SDSS\,J1617+1620, demonstrating that observations of dusty and
gaseous discs around metal-polluted white dwarfs  can provide further insight into the
dynamics of such systems. This will also allow us explore the
nature and composition of the accreted material in greater detail and
better understand the post main sequence evolution of extrasolar
planetary systems. Continued observations of SDSS\,J1617+1620 and other
white dwarfs with gaseous discs will be required to further investigate the highly dynamical nature of evolved planetary systems. 

\section{Acknowledgements}
The research leading to these results has received funding from the
European Research Council under the European Union's Seventh Framework
Programme (FP/2007-2013) / ERC Grant Agreement n. 320964 (WDTracer).
BTG was supported in part by the UK’s Science and Technology
Facilities Council (ST/I001719/1). DJW would like to thank M
Hollands for help with Figure (\ref{fig:specfit}).
 
This paper has made use of observations from the SDSS-III, funding for which has been provided by the Alfred P. Sloan Foundation, the Participating Institutions, the National Science Foundation, and the U.S. Department of Energy Office of Science. 
It is also based on observations made with the William Herschel Telescope on the island of La Palma by the Isaac Newton Group in the Spanish Observatorio del Roque de los Muchachos of the Instituto de Astrofísica de Canarias, spectra obtained with the Gemini North Telescope under program GN-2010A-Q-94, and observations made with ESO Telescopes at
the La Silla Paranal Observatory under programme ID 093.D-0838(A),087.D-0139(C),091.D-0296(A),and 386.C-0218(E).

\bibliographystyle{mn_new.bst}
\bibliography{aamnem99,aabib}

\bsp

\end{document}